\documentclass[prl,twocolumn,showpacs,preprintnumbers,amsmath,amssymb,a4paper]{revtex4-1}
\usepackage{graphicx}
\begin{document}

\title{Topological insight into the non-Arrhenius mode hopping of semiconductor ring lasers}

\author{S.~Beri$^{1}$, L.~Gelens$^1$, M.~Mestre$^1$, G.~Van~der~Sande,$^1$ G.~Verschaffelt$^1$, A.~Scir\`e$^2$, G.~Mezosi$^3$, M.~Sorel$^3$ and J.~Danckaert$^{1}$}

\affiliation{
$^1$Department of Applied Physics and Photonics, Vrije Universiteit Brussel,
Pleinlaan 2, 1050 Brussels, Belgium;\\
$^2$Instituto de F\'isica Interdisciplinar y Sistemas Complejos 
(IFISC, CSIC-UIB), Campus Universitat Illes Balears, E-07122 Palma de Mallorca, Spain;\\
$^3$Department of Electronics \& Electrical Engineering, University of Glasgow, Rankine Building, Oakfield Avenue,
Glasgow, G12 8LT, United Kingdom.
}

\date{\today}


\begin{abstract} 
We investigate both theoretically and experimentally the stochastic switching between two counter-propagating lasing modes of a semiconductor ring laser.
Experimentally, the residence time distribution cannot be described by a simple one parameter Arrhenius exponential law and reveals the presence of two different mode-hop scenarios with distinct time scales. In order to elucidate the origin of these two time scales, we propose a topological approach based on a two-dimensional dynamical system. 
\end{abstract}
\maketitle

Fluctuations in active optical systems such as lasers is one of today's technological challenges as well as a fundamental problem of modern physics as they are the result of the quantum nature of the interaction between light and matter \cite{Loudon}.
Fluctuations are e.g. responsible for longitudinal mode switching in semiconductor lasers \cite{Ohtsu86a}, polarization mode-hopping in Vertical Cavity Surface Emitting Lasers (VCSELs) \cite{Giacomelli98a,Willemsen00a,Nagler03a}, and they play a fundamental role in stochastic and coherence resonances of optical systems \cite{Giacomelli00a,Fioretti93a,McNamara88a}.

Semiconductor ring lasers (SRLs) are a particular class of lasers whose operation is strongly affected by stochastic fluctuations.
The circular geometry of the active cavity allows a SRL to operate in two possible directions, namely clockwise mode ($CW$) and counter-clockwise mode ($CCW$). 
From the application point of view, SRLs are ideal candidates for all-optical information-storage. \cite{SorelOL2002,LiangAPL1997,HillNature2004}. 
From a theoretical point of view, SRLs represent the optical prototype of nonlinear $Z_2$-symmetric systems \cite{Kuznetsovbook}, which appear in many fields of physics. 

Fluctuations induce spontaneous abrupt changes in the SRL's directional operation from $CW$ to $CCW$ and vice versa, and therefore represent a major limitation to their successful applications for instance as optical memories. An in-depth understanding of the mode-hopping in SRLs would shed light on the stochastic properties of the large class of $Z_2$-symmetric systems. 
In spite of its importance, the problem of spontaneous directional switches in SRLs remains unaddressed, partly due to the high dimensionality of the models that have been proposed for SRLs \cite{SorelOL2002,Zeglache88a}.

In this paper, we address the problem of such fluctuations both theoretically and experimentally. We experimentally investigate the properties of the residence time distribution (RTD) that quantifies the mode hopping.
Our theoretical analysis is based on an asymptotic reduction of a full rate-equation model to a $Z_2$-symmetric planar system \cite{VanderSandeJPhysB2008}.

\begin{figure}[]
\centering
\includegraphics[width=6cm]{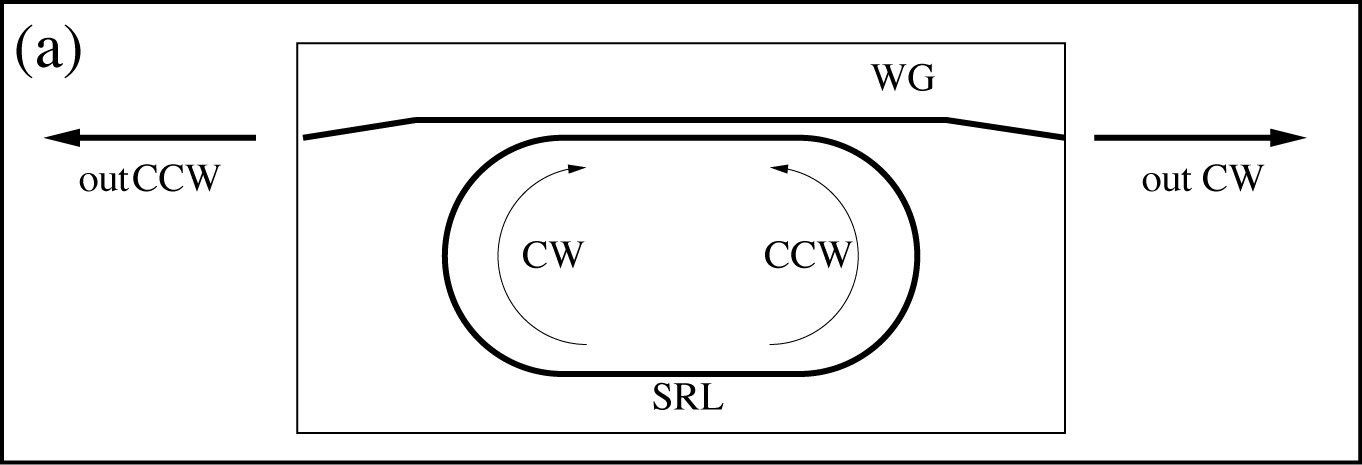}
\includegraphics[width=8.4cm]{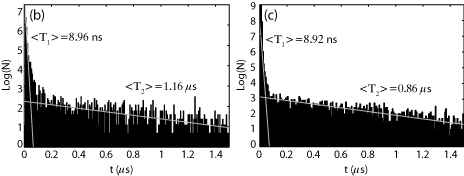}
\caption{\label{Fig:RTD} (a) Schematic sketch of the set-up. (b) Measured RTD in the $CCW$ mode for a laser current of $J_{ring}=39.92$mA; the waveguide is biased at $8.22$mA and the temperature is $21.55^\circ$C.
(c) RTD obtained by simulating Eqs.~(\ref{Eq::Field1::Original})-(\ref{Eq::Carriers::Original}) with the following parameters: $\mu = 1.59$, $\alpha=3.5$, $s=0.005$, $c=0.01$, $k=0.44$ns$^{-1}$, $\phi_{k}=1.5$, $D=6.5\cdot 10^{-5}$ ns$^{-1}$. Best fitting lines are shown in white.}
\end{figure}

We consider here an InP-based multiquantum-well SRL with a racetrack geometry and a free-spectral-range of $53.6$ GHz. The device operates in a single-transverse, single-longitudinal mode regime at wavelength $\lambda = 1.56 \mu$m.
However, it will be clear from the rest of the discussion that our analysis is general and applies to any kind of ring geometry.
A wave\-guide has been integrated on the same chip in order to couple power out from the ring. This bus waveguide can be independently biased in order to reduce absorption losses.
The waveguide crosses the facets of the chip under a $10^\circ$ angle in order to minimize back-reflections.
The chip is mounted on a copper mount and thermally controlled by a Peltier element which is stabilized with an accuracy of $0.01^\circ$C.
The power emitted from the chip is coupled to a multimode fibre and detected with a $2.4$GHz photodiode connected to an oscilloscope.
We forward-bias the waveguide at a fixed current of $8.22 \pm 0.01$ mA in order to achieve transparency, and we bias the device with increasing DC current until it reaches the threshold at $\approx 31.5$ mA. The stochastic mode-hopping starts at approximately $39$ mA.
In the mode-hopping region, the time series of the power emitted in the $CCW$ mode have been digitally recorded with the oscilloscope for different values of the bias current.
Due to the anti-correlation of the two counterpropagating modes \cite{SorelOL2002}, a drop in power in the $CW$ mode corresponds to an increase in power emitted in the $CCW$ mode.
The $CCW$ mode's RTD has been obtained from the time series and is plotted in Fig.~\ref{Fig:RTD}(b).
It is evident from Fig.~\ref{Fig:RTD}(b) that two well-separated time scales are present in the RTD. The white lines depict the best linear fit in this logarithmic plot of the regions where the residence times are exponentially distributed with one of both characteristic time scales [$\propto \exp{(-t / \langle T_{1,2} \rangle)}$]. The fast time scale corresponds to an average residence time $\langle T_1 \rangle$ of 10ns, whereas the slow time scale events have an average residence time $\langle T_2 \rangle$ of approximately 1$\mu$s.
An increase in the bias current affects the time scales in different ways: $\langle T_1 \rangle$ remains almost constant, whereas $\langle T_2 \rangle$ increases with bias current.
The appearence of the short time scale is a characteristic feature of SRLs: similar experiments performed in VCSELs \cite{Willemsen00a,Nagler03a} or in Dye ring lasers \cite{Lett85a} lead to RTDs that are well fitted by a one-parameter Arrhenius distribution.
A spectral analysis of the time series reveals a relaxation-oscillation (RO) frequency between 1.0 and 2.0 GHz and therefore rules out ROs as the origin of the non-Arrhenius features of RTD.

\begin{figure*}[t!]
\centering
\includegraphics[width=13cm]{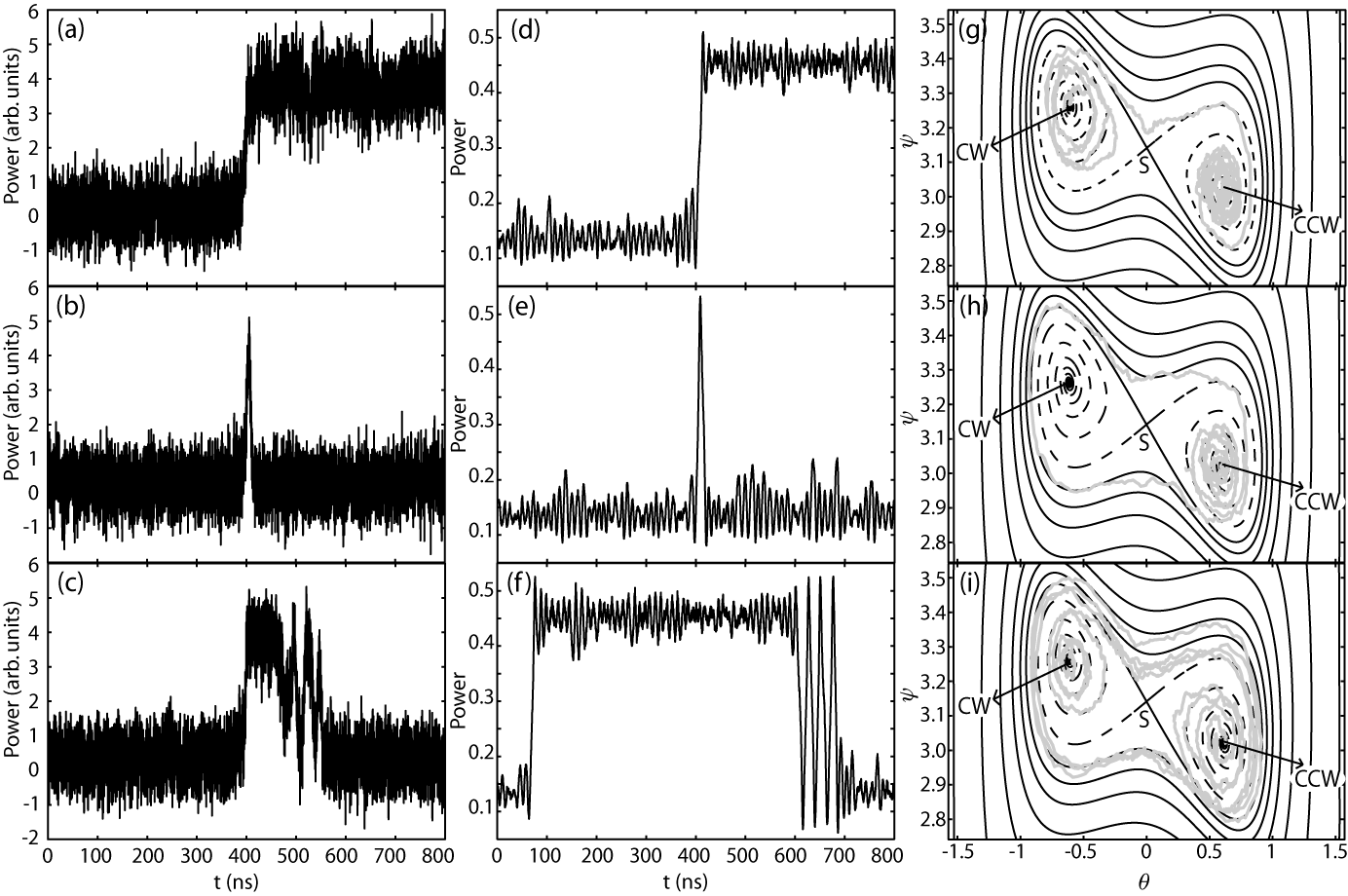}
\caption{\label{Fig:TS} 
Comparison between measured time series (a)-(c), simulated time series (d)-(f) and phase space trajectories (g)-(i) for different kind of transitions.
In the experiment the ring was biased at $J_{ring}=39.86$mA; the waveguide was biased at $8.22$mA and the temperature was $21.55^\circ$C.
In the numerical simulation the following parameters were used: $\mu = 1.59$, $\alpha=3.5$, $s=0.005$, $c=0.01$ $k=0.44$ns$^{-1}$, $\phi_{k}=1.5$, $D=6.5\cdot 10^{-5}$ ns$^{-1}$.
The notation of (g)-(i) is as follow: $CW$, $CCW$ are the stable states, $S$ is the saddle; solid line: stable manifold of $S$, dashed line: unstable manifold of $S$, gray: projection of the time series (d)-(f).}
\end{figure*}

In order to clarify the origin of the two separate time scales, we have directly investigated the recorded time traces. Qualitatively different switches are revealed in these time traces as shown in Fig.~\ref{Fig:TS}(a)-(c).
Fig.~\ref{Fig:TS}(a) shows the system leaving the initially lasing $CW$ mode and settling into the $CCW$ mode.
Short excursions of the duration of approximately 10ns to the opposite mode have also been observed as shown in Fig.~\ref{Fig:TS}(b). 
Opposite to the case of Fig.~\ref{Fig:TS}(a), during these short excursions, the system does not settle in the other mode.
The durations of such events match the fast time scale $\langle T_1 \rangle$  in the RTD.
Instead, the second time scale in the RTD is related to long residences in one mode such as depicted in Fig.~\ref{Fig:TS}(a). The long residences are distributed in an exponential way with a mean-residence-time of $1.16\mu$s.
We would like to stress here that the RTD is not a simple superposition of independent events such as those in Fig.~\ref{Fig:TS}(a),(b), but that more complex transitions are also present in the system, such as the one depicted in Fig.~\ref{Fig:TS}(c). In the latter, a number of consecutive, correlated short time excursions are performed before the system settles in the counterpropagating mode.

In order to explain our experimental results, we start from a classical rate-equation model, which accounts for both saturation and backscattering effects \cite{SorelOL2002}. 
The evolution of the slowly varying envelopes of both complex counter-propagating fields ($E_1$ and $E_2$) and the carrier density ($N$) is described by the following set of equations:
\begin{eqnarray}
\dot{E}_{1,2} &=& \kappa(1+i\alpha)\left[N\left(1-s|E_{1,2}|^2-c|E_{2,1}|^2\right)-1\right]E_{1,2} \nonumber \\
&&- k e^{i \phi_k}E_{2,1}, \label{Eq::Field1::Original}\\
\dot{N} &=& \gamma [ \mu -N - N\left(1-s|E_1|^2-c|E_2|^2\right)|E_1|^2 \nonumber \\
&&- N\left(1-s|E_2|^2-c|E_1|^2\right)|E_2|^2] , \label{Eq::Carriers::Original}
\end{eqnarray}
where the electric fields and the carrier inversion have been non-dimensionalized and the dot represents differentiation with respect to time.
In Eqs.~(\ref{Eq::Field1::Original})-(\ref{Eq::Carriers::Original}), $\kappa$ is  the field decay rate, and $\gamma$ is the decay rate of the carrier population. $\alpha$ is the linewidth enhancement factor, $\mu$ is the renormalized injection current ($\mu\approx0$ at transparency, $\mu\approx1$ at lasing threshold). Self- and cross-saturation effects are the result of spatial and spectral hole burning in the system and are described by $s$ and $c$. 
The backscattering (for instance due to defects or backreflections \cite{Spreeuw90}) is characterised by a certain amplitude $k$ and phase $\phi_k$.
Although perfectly adequate to model the SRL operation, nevertheless such five-dimensional model needs to be simplified in order to allow for a clear topological analysis.
A reduced two-dimensional model has been proposed \cite{VanderSandeJPhysB2008}, which describes the asymptotic behaviour of the ring laser on time scales that are slower than the RO:
\begin{eqnarray}
\dot{\theta} &= -2\sin\phi_k\sin\psi + 2 \cos\phi_k\cos\psi\sin\theta \nonumber \\
& + J \sin\theta\cos\theta , \label{Eq::Theta}\\
\cos\theta \dot{\psi} &= \alpha J \sin\theta\cos\theta + 2 \cos\phi_k \sin\psi \nonumber \\
&+ 2 \sin\phi_k \cos\psi \sin\theta. \label{Eq::Psi}
\end{eqnarray}
where $\theta = 2 \arctan \sqrt{|E_2|^2 / |E_1|^2} - \pi/2 \in [-\pi/2,\pi/2]$ 
represents the relative modal intensity and $\psi \in [0,2\pi]$ is the phase difference between the counter-propagating modes. The pump current has been rescaled
as $J=\kappa (c-s)(\mu-1)/k$.  The use of Eqs.~(\ref{Eq::Theta})-(\ref{Eq::Psi}) 
is justified by the fact that all the observed time scales [Fig.~\ref{Fig:RTD}(b)] are slower than the relaxation oscillations.

When the SRL operates in a unidirectional regime, the phase portrait of the system is exemplified in Figure \ref{Fig:TS}(g)-(i).
Four stationary solutions exist for Eqs.~(\ref{Eq::Theta})-(\ref{Eq::Psi}): an unstable in-phase bidirectional state in $(0,0)$ (not shown); two symmetric stable states $CW$ and $CCW$ at 
$\psi \approx \pi$, both corresponding to unidirectional operation; and a saddle point $S$
 in $(0 , \pi)$ which is the unstable out-of-phase bidirectional solution. 
 The stable manifold of $S$ separates the basins of attractions of $CW$ and
$CCW$, whereas the unstable manifold connects $S$ with $CW$ and $CCW$. At a critical value $J_{hom}$ of the current, a homoclinic bifurcation takes place in the system and the stable and unstable manifold of $S$ coincide \cite{VanderSandeJPhysB2008}. When $J>J_{hom}$ the stable manifold spirals around $S$ and the basins of attraction of $CW$ and $CCW$ fold around each-other.
Increasing the parameter $J$ continuously unfolds the stable manifold.

When noise is present in the system, a rare, large fluctuation may drag the system away from an initial stationary state to the basin of attraction of the opposite mode.
It is known from the theory of stochastic transitions in nonlinear systems that in the limit of small noise strength a transition takes place in a ballistic way along a \emph{most probable escape path} (MPEP) which can be calculated by solving an auxiliary Hamiltonian system \cite{Dykman92a}.
As our arguments rely only on the topological features of the system, an exact calculation of the MPEP is not required.
In a system such as Eqs.~(\ref{Eq::Theta})-(\ref{Eq::Psi}), it is known that the MPEP connects a stationary state with the saddle $S$ \cite{Silchenko03a}. More specifically, the MPEP approaches $S$ along its stable manifold. 
Once the saddle is reached, the transition to the opposite equilibrium state is completed by following the unstable manifold of the saddle.
However, when the system's parameters are close to the homoclinic bifurcation,
the folds of the stable manifold cluster very tightly around each other, and the stable and unstable manifolds of the saddle become very close.
Therefore, even at very low noise intensities, the system can diffuse between the folds of the stable manifold or between the stable and the unstable manifold.
Such topological considerations can explain the separated time scales $\langle T_{1,2} \rangle$ as well as the features of the transitions shown in Fig.~\ref{Fig:TS}(a)-(c). A noise induced activation is responsible for the slow time scale $\langle T_{1} \rangle$, whereas a noise-sustained rotation along the folds of the stable manifold of $S$ produces the short excursion shown in Fig.~\ref{Fig:TS}(b). 

We stress that SRLs are out-of-equilibrium systems \cite{Melnikov91a}; although the phase portrait of the system is similar to the one obtained in the case of an underdamped Duffing oscillator,
the topology of the manifold is related to the vicinity of the homoclinic bifurcation, but cannot be considered as a sign of weak dissipation.

In order to confirm our topological arguments, we perform numerical simulations of Eqs.~(\ref{Eq::Field1::Original})-(\ref{Eq::Carriers::Original}) using relevant parameters \cite{SorelOL2002}.
The spontaneous emission noise has been introduced in Eq.\ (\ref{Eq::Field1::Original}) phenomenologically as complex uncorrelated zero-mean stochastic terms described by the correlation terms:  
$\langle \xi_{i}\left( t+\tau \right) \xi^*_{j}\left( t \right) \rangle = 2D \delta_{ij} \delta \left( \tau \right)$ where $i,j = 1,2$ and $D$ is the noise intensity. Carrier noise has been disregarded as its relevance in directional mode-hopping was proven negligible \cite{Lett85a}.

Three examples of numerically obtained stochastic transitions are reported in Fig.~\ref{Fig:TS}(d)-(f) and compared with the experimental  ones \footnote{One should be aware that the RMS for the noise background observed in the experimental time series is dominated by the electrical noise floor of the oscilloscope and the detector.}.
The same trajectories projected in the $\theta-\psi$ plane are shown in Fig.~\ref{Fig:TS}(g)-(i) together with the invariant manifolds.
The trajectory in Fig.~\ref{Fig:TS}(d) shows a simple mode-hop as in Fig.~\ref{Fig:TS}(a).
In phase space [Fig.~\ref{Fig:TS}(g)] such trajectory corresponds to a stochastic path which approaches the saddle and relaxes to the opposite state along the unstable manifold of $S$. 
A short excursion to the opposite direction is shown in Fig.~\ref{Fig:TS}(e), corresponding to the experimental time trace shown in Fig.~\ref{Fig:TS}(b).
In phase space [Fig.~\ref{Fig:TS}(h)] the trajectory rotates around the saddle, but remains inside the basin of attraction of the initial mode.
In Fig.~\ref{Fig:TS}(f), we show an example of a more complicated transition similar to the experimental trajectory shown in Fig.~\ref{Fig:TS}(c).
In phase space [Fig.~\ref{Fig:TS}(i)] the system rotates three times around $S$ before settling in the opposite state.
The simulations shown in Fig.~\ref{Fig:TS}(d)-(f) are in good qualitative agreement with the experiments in Fig.~\ref{Fig:TS}(a)-(c). A direct simulation of the reduced model Eqs.~(\ref{Eq::Theta})-(\ref{Eq::Psi}) leads to similar results.

We remark that the described effect takes place in regions of the parameter space $\left( J , \alpha, \phi_k \right)$ which are close to the homoclinic bifurcation. In the unphysical case of $k=0$ (no backscattering) the manifolds are completely unfolded for every value of $\phi_k$ and $J$ and the effect disappears. 

Finally, we have extracted the RTD from the simulated time series [Fig.~\ref{Fig:RTD}(c)]. The simulations show the same non-Arrhenius structure as in the experimental RTD, and quantitatively reproduce the time scales $\langle T_{1,2} \rangle$.

In conclusion, we have shown that the RTD for the mode-hopping in SRLs is strongly non-Arrhenius and cannot be described by a single transition rate.
Two separate time scales are observed in the RTD.
Our explanation of this behaviour is based on topological properties of the phase space of the asymptotically reduced model  [Eqs.~(\ref{Eq::Theta})-(\ref{Eq::Psi})] in the vicinity of a homoclinic bifurcation. 
We have found that during a noise induced transition, the system explores the region of the phase space where the stable manifold of the saddle $S$ rotates around the saddle. Even at low noise intensity, the system can diffuse between different folds of the manifold  leading
to different kind of transitions such as those experimentally observed in Fig.~\ref{Fig:TS}(a)-(c).
Numerical simulations agree well with the experiments. The simulated RTD is in good quantitative agreement with the experimental one.
We finally remark that the topology of the manifolds that we have observed for the planar system [Eqs.~(\ref{Eq::Theta})-(\ref{Eq::Psi})] has been observed in other optical system such as in the Ikeda map applied to a nonlinear ring resonator \cite{Hammel85a}, and it is a general feature of two-dimensional dynamical systems which are $Z_2$-invariant \cite{Kuznetsovbook}. Therefore, SRLs are excellent optical prototypes  for the experimental exploration of stochastic systems with $Z_2$-symmetry.

This work has been partially funded by the European Union under project IST-2005-34743 (IOLOS). This work was supported by the Belgian Science Policy Office under grant No.\ IAP-VI10. G.V. is a Postdoctoral Fellow and LG  is a PhD Fellow of the Research Foundation - Flanders (FWO).


\begin{thebibliography}{99}

\bibitem{Loudon}
R.~Loudon, ``The Quantum Theory of Light'', (Oxford University Press, Oxford, 2000)

\bibitem{Ohtsu86a}
M.~Ohtsu, Y. Teramachi, Y.~Otsuka, and A.~Osaki,
IEEE J. Quantum Electron. {\bf 2239,}, 535 (1986)

\bibitem{Giacomelli98a}
G.~Giacomelli, and F.~Marin,
Quantum Semiclass. Opt {\bf 10}, 469 (1998)

\bibitem{Willemsen00a}
M. B. Willemsen, M. P. van Exter, and J. P. Woerdman, Phys. Rev. Lett. 84, 4337 (2000) 

\bibitem{Nagler03a}
B.~Nagler, M.~Peeters, J.~Albert, G.~Verschaffelt, K.~Panajotov , H.~Thienpont, I.~Veretennicoff, J.~Danckaert,  S.~Barbay, G.~Giacomelli, and F.~Marin,
Phys. Rev. A {\bf 68}, 013813 (2003)

\bibitem{Giacomelli00a}
G.~Giacomelli,
M.~Giudici, S.~ Balle,
and J.~ R.~ Tredicce,
Phys. Rev. Lett. {\bf 84}, 3298 (2000)

\bibitem{Fioretti93a}
A.~Fioretti, L.~Guidoni, R.~Mannella and E.~Arimondo,
Jour. Stat. Phys. {\bf 70}, 403 (1993)

\bibitem{McNamara88a}
B. McNamara, K.~Wiesenfeld and R.~Roy,
Phys. Rev. Lett. {\bf 60}, 2626 (1988)

\bibitem{SorelOL2002} M. Sorel, J. P. R. Laybourn, A. Scir{\'e}, S. Balle, G. Giuliani, R. Miglierina, and S. Donati,  Opt. Lett. {\bf 27}, 1992 (2002).

\bibitem{LiangAPL1997} J. J. Liang, S. T. Lau, M. H. Leary, and J. M. Ballantyne,  Appl. Phys. Lett. {\bf 70}, 1192 (1997).

\bibitem{HillNature2004} M. T. Hill, H. J. S. Dorren, T. de Vries, X. J. M. Leijtens, J. H. den Besten, B. Smalbrugge, Y. S Oei, H. Binsma, G. D. Khoe, and M. K. Smit,  Nature {\bf 432}, 206 (2004).
\bibitem{Kuznetsovbook} A. {Kuznetsov},  Springer Book, 3rd edition (2004).

\bibitem{Zeglache88a}
H.~Zeghlache, P~Mandel, N.~B.~Abraham, L.~M.~Hoffer, G.~L.~Lippi, and T.~Mello 
Phys. Rev. A, \textbf{37}, 470 (1988).

\bibitem{VanderSandeJPhysB2008} G. {Van der Sande}, L. Gelens, P. Tassin, A. Scir{\'e} and J. Danckaert,  J.Phys.B {\bf 41}, 095402 (2008).

\bibitem{Spreeuw90}
R.~ J.~ C.~ Spreeuw, R.~ C.~ Neelen, N.~ J.~ van Druten, E.~ R.~ Eliel, and J.~ P.~ Woerdman,  Phys. Rev. A,
\textbf{42}, 4315 (1990).

\bibitem{Lett85a}
P.~Lett and L.~Mandel J. Opt. Soc. Am. B {\bf 2}, 1615 (1985);

\bibitem{Dykman92a}
M. I. Dykman, P. V. E. McClintock, V. N. Smelyanski, N. D. Stein and N. G. Stocks, Phys. Rev. Lett. {\bf 68}, 2718 - 2721 (1992)

\bibitem{Silchenko03a}
A.~N.~Silchenko, S.~Beri, D.~G.~ Luchinsky, P.~V.~E.~ McClintock, Phys. Rev. Lett. 91, 174104 (2003) 

\bibitem{Melnikov91a}
V.~I.~Mel’nikov, Phys. Rep. {\bf 209}, 1--71 (1991)

\bibitem{Hammel85a}
S.~M.~Hammel, C.~Jones, J.~V.~Moloney, J.~Opt.~Soc.~Am.~B {\bf 2}, 552--564 (1985)

\end{thebibliography}
\end{document}